\documentclass{elsart}

\usepackage{graphicx}

\usepackage{amssymb}

\begin{document}

\begin{frontmatter}



\title{Electron impact excitations of S$_{2}$ molecules}


\author[fax]{Motomichi Tashiro}
\ead{tashiro@fukui.kyoto-u.ac.jp}
\thanks[fax]{FAX: +81 75 781 4757}
\address{
    Fukui Institute for Fundamental Chemistry,
    Kyoto University,
    Takano-Nishihiraki-cho 34-4, Sakyo-ku, 
    Kyoto 606-8103, Japan
    }

\begin{abstract}
Low-energy electron impact excitations of S$_2$ molecules are 
studied using the fixed-nuclei R-matrix method based on 
state-averaged complete active space SCF orbitals. 
Integral cross sections are calculated for elastic electron collision 
as well as impact excitation of the 7 lowest excited electronic states. 
Also, differential cross sections are obtained for elastic collision 
and excitation of the ${a}^1\Delta_{g}$, ${b}^1\Sigma_{g}^{+}$ and 
${B}^3\Sigma_{u}^{-}$ states. 
The integrated cross section of optically allowed excitation of the 
${B}^3\Sigma_{u}^{-}$ state agrees reasonably well with the available 
theoretical result. 
\end{abstract}

\end{frontmatter}

\section{Introduction}
\label{1}

%
%
%
%
Diatomic sulfur, S$_2$, has been subject of many theoretical and 
spectroscopic investigations for long time \cite{S2spectr,Swope79,Saxon80}. 
We can find S$_2$ molecules at various natural and industrial plasmas 
containing sulfur compounds. For example, emissions and absorptions of 
S$_2$ molecules have been observed in the atmospheres of 
Jupiter \cite{S2Jupiter} and its satellite Io \cite{S2Io}. 
They are also observed in the atmospheres of some comets \cite{S2Hyakutake}.
In industrial condition, S$_2$ molecules can be seen in reactive ion 
etching process using SF$_6$ molecules \cite{S2etching}. 
Sulfur lamps contain S$_2$ molecules as an important ingredient \cite{S2Lamp}. 
Although electron collision with S$_2$ molecules is an important elementary 
process in these plasmas, there has been little work on this subject. 
As far as we are aware, no experimental measurement nor theoretical 
calculation of electron-S$_2$ elastic cross section have been performed. 
Garrett et al. \cite{S2IP} calculated integral cross section for electron 
impact excitation of the 
S$_2$ B$^3 \Sigma_u^-$, 2$^3 \Sigma_u^-$ and B''$^3 \Pi_u$ states using 
the semiclassical impact-parameter (IP) method extended to include 
nuclear motion. 
Since the IP method is designed to treat optically allowed transitions, 
electron impact excitation to the other electronic state was not 
investigated. 
Le Coat et al. \cite{S2DAexp} performed experimental measurement on dissociative electron 
attachment of S$_2$ molecules and identified two resonances, however, 
they did not present absolute value of the cross section. 

Recently, we have performed the R-matrix calculations on the spin-exchange 
effect in electron collision with homo-nuclear open-shell diatomic molecules 
including S$_2$ \cite{Tashiro2008}. We have also calculated the integral 
cross section of elastic electron collision with S$_2$ molecules 
for the first time. 
Here, we extend our previous study to the electron 
impact excitations of S$_2$ molecules. 
Since S$_2$ molecule has a valence electron structure similar to O$_2$, 
comparison of the cross sections will be interesting. 
Also, the cross section of electron impact excitation of the 
S$_2$ B$^3 \Sigma_u^-$ would be important in analyzing sulfur plasma. 
As in our previous works on electron impact excitation of O$_2$ and 
N$_2$ \cite{tashiroO2ics,tashiroO2DCS,tashiroN2}, 
we employ the fixed-nuclei R-matrix method using state-averaged complete 
active space SCF molecular orbitals. 
The size of basis set for the scattering electron is slightly extended 
in this work compared to our previous calculation on the spin-exchange 
effect \cite{Tashiro2008}.

\section{Theoretical methods}
\label{2}

The details of the R-matrix method has been described in the literature 
\cite{Mo98,Bu05,Go05}, thus we do not repeat general explanation of 
the method here. 
We used a modified version of the polyatomic programs in the UK molecular 
R-matrix codes \cite{Mo98} in this work. 
These programs utilize the Gaussian type orbitals (GTO) to 
represent target electronic states as well as a scattering electron. 
In the present R-matrix calculations, we have included 13 target states; 
${X}^3\Sigma_{g}^{-}$, ${a}^1\Delta_{g}$, ${b}^1\Sigma_{g}^{+}$, 
${c}^1\Sigma_{u}^{-}$, ${A'}^3\Delta_{u}$, ${A}^3\Sigma_{u}^{+}$,
${B'}^3\Pi_{g}$, ${B}^3\Sigma_{u}^{-}$, ${1}^1\Pi_{g}$, 
${1}^1\Delta_{u}$, ${B''}^3\Pi_{u}$, ${1}^1\Sigma_{u}^{+}$, 
and ${1}^1\Pi_{u}$. 
These target states were represented by valence configuration interaction 
wave functions constructed by state averaged complete active space SCF 
(SA-CASSCF) orbitals. 
In this study, the SA-CASSCF orbitals were obtained by calculations with 
MOLPRO suites of programs \cite{molpro}. 
The target orbitals were constructed from the cc-pVTZ basis 
set\cite{1993JChPh..98.1358W}. 
Fixed-nuclei approximation was employed with inter-nuclear distance of 
3.7 a$_0$. 
The radius of the R-matrix sphere was chosen to be 13 a$_0$ in our  
calculations.
In order to represent the scattering electron, we included diffuse
Gaussian functions up to $l$ = 5, with 13 functions for $l$ = 0, 
11 functions for $l$ = 1, 10 functions for $l$ = 2, 
8 functions for $l$ = 3, 6 functions for $l$ = 4 and 
5 functions for $l$ = 5. 
Exponents of these diffuse Gaussians were taken from Faure 
et al. \cite{Fa02}. 
In addition to these continuum orbitals, we included 8 extra virtual 
orbitals, one for each symmetry. 
The construction of the configuration state functions (CSFs) for 
the electron-molecule system is the same as in our previous paper\cite{Tashiro2008}. 
Note that 
The R-matrix calculations were performed for all 8 irreducible 
representations of the D$_{2h}$ symmetry, 
$A_g$, $B_{2u}$, $B_{3u}$, $B_{1g}$, $B_{1u}$, $B_{3g}$, $B_{2g}$ 
and $A_u$, in doublet and quartet spin multiplicities 
of the whole system.

One of the transitions studied in this work, the excitation of the 
${B}^3\Sigma_{u}^{-}$ state from the ground state, is optically allowed. 
Thus, we have to consider the effect of transition dipole moment between 
these two states. 
A lot of $l$ partial-waves has to be included in the R-matrix calculation 
to obtain converged cross sections because of the long-range interaction of the dipole, 
although it is difficult to include partial waves with $l \geq 7$ in 
the usual ab initio R-matrix calculation. 
In this work, the R-matrix calculations are performed with partial waves 
up to $l = 5$. 
The effects of the higher $l$ partial waves are included 
by the Born closure approximation as in the previous works \cite{Crawford1971,Gibson87}. 
Following Gibson et al.\cite{Gibson87}, we evaluate 
the differential cross sections (DCSs) with the Born correction, 
${d \sigma^{\rm BC} / d \Omega}$, by the expression, 
\begin{equation}
\frac{d \sigma^{\rm BC}}{d \Omega} = \frac{d \sigma^{\rm FBA}}{d \Omega} + 
\left[
\frac{d \sigma^{\rm R-matrix}}{d \Omega} - \frac{d \sigma^{\rm FBA}_{\rm FE}}{d \Omega}
\right].
\end{equation}
Here, ${d \sigma^{\rm FBA}/d \Omega}$ is the DCS obtained by the first Born 
approximation, ${d \sigma^{\rm R-matrix}/d \Omega}$ is the cross section obtained by 
the R-matrix calculation and ${d \sigma^{\rm FBA}_{FE}/d \Omega}$ is the DCS 
from the first Born approximation including the same number of partial waves as in 
the R-matrix calculation. ${d \sigma^{\rm FBA}_{FE}/d \Omega}$ is evaluated by 
the angular momentum representation of the T-matrix elements 
for the first Born approximation. These T-matrix elements as well as 
${d \sigma^{\rm FBA}/d \Omega}$ are available in close form \cite{Itikawa1969}. 
The total cross sections are obtained by the integration of eq.(1).

\section{Results and Discussion}
\label{3}

In table \ref{tab1}, excitation energies of S$_2$ molecule 
obtained from the CASSCF calculation in this work 
are compared with MRD CI vertical excitation energies of Hess et 
al.\cite{ChemPhys.71.79}, MRCI adiabatic excitation energies of Kiljunen 
et al.\cite{2000JChPh.112.7475K} and experimental values quoted in 
Hess et al.  
Our results agree well with the previous calculations and experimental results 
for the lowest two excitations. 
For excitations to the higher electronic states, deviations in excitation 
energies become larger partly because of difference of 
adiabatic and vertical excitation energy. 

\begin{figure}
\includegraphics{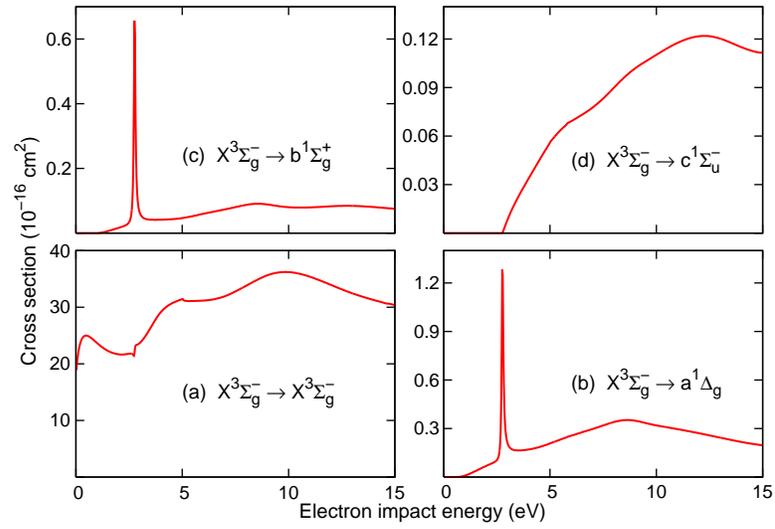}%
 \caption{\label{fig1} 
  Integral elastic (panel a) and excitation cross sections of the 
  $a^{1} \Delta_{g}$ (panel b), $b^{1} \Sigma_{g}^{+}$ (panel c) 
  and ${c}^{1} \Sigma_{u}^{-}$ (panel d) states. 
   }
\end{figure}

In figure \ref{fig1}, the integral cross sections (ICSs) are shown for 
electron S$_2$ elastic collision and excitations of the 
${a}^1\Delta_{g}$, ${b}^1\Sigma_{g}^{+}$ and ${c}^1\Sigma_{u}^{-}$ states. 
The magnitude of the elastic ICS is about 20 $\times 10^{-16} {\rm cm}^2$ in 
low energy region below 3 eV, then it increases to be 30 $\times 10^{-16} 
{\rm cm}^2$ at energies above 5 eV. 
The ICS of the excitation to the ${a}^1\Delta_{g}$ state increases gradually 
from threshold to 8.5 eV, where it takes a maximum value of 
0.35 $\times 10^{-16} {\rm cm}^2$, then it decreases again. 
The ICSs of the excitation to the ${b}^1\Sigma_{g}^{+}$ and 
${c}^1\Sigma_{u}^{-}$ states also increase gradually from threshold. 
In both cases, the maximum value of the ICS is about 0.1 
$\times 10^{-16} {\rm cm}^2$. 
Around 2.7 eV, a sharp resonance peak with width 0.08 eV exists in the ICSs of the 
${a}^1\Delta_{g}$ and ${b}^1\Sigma_{g}^{+}$ excitations. 
Also, a kink structure is observed in the elastic ICS at the same energy. 
All of these structure belongs to the ${}^2 \Pi_u$ symmetry partial cross 
sections. 
We analyzed the CSFs and found that the kink and peaks at 2.7 eV are likely 
related to a resonance with configuration 
$({\rm core})^{20}(4\sigma_g)^2(4\sigma_u)^2(5\sigma_g)^2(2\pi_u)^3(2\pi_g)^4$, 
which is obtained from an attachment of the scattering electron to the excited 
${c}^1 \Sigma_u^-$, ${A'}^3 \Delta_u$ and ${A}^3 \Sigma_u^+$ states of 
S$_2$ with configuration 
$({\rm core})^{20}(4\sigma_g)^2(4\sigma_u)^2(5\sigma_g)^2(2\pi_u)^3(2\pi_g)^3$. 
The width of this resonance is about 0.08 eV at R = 3.7 a$_0$.   
We checked the behaviour of this resonance as a function of bond-length, 
and found that it approaches the atomic limit ${\rm S}({}^3P)$ +${\rm S}^-({}^2P)$.
Since the location of this resonance is higher than the atomic limit, 
dissociative electron attachment may occur through this resonance. 

\begin{figure}
\includegraphics{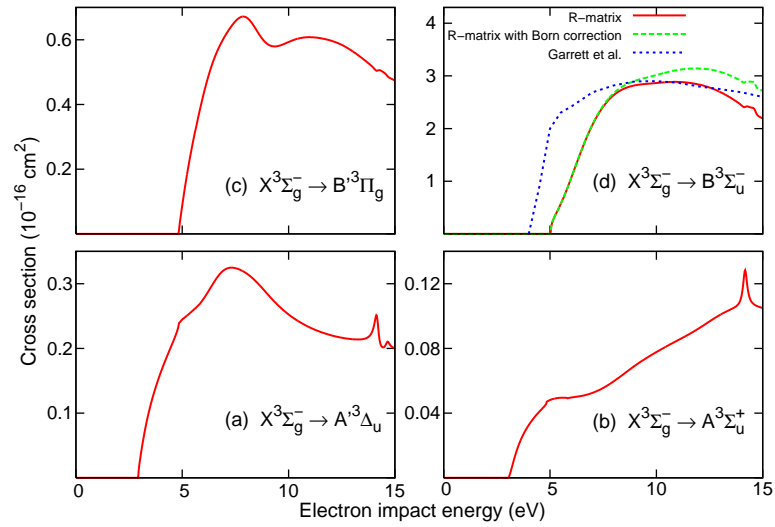}%
 \caption{\label{fig2} 
  Integral excitation cross sections of the 
  ${A'}^{3} \Delta_{u}$ (panel a), $A^{3} \Sigma_{u}^{+}$ (panel b),  
  ${B'}^{3} \Pi_{g}$ (panel c) and ${B}^{3} \Sigma_{u}^{-}$ (panel d) states. 
  The R-matrix cross section with Born correction and 
  the previous results of Garrett et al. \cite{S2IP} are also included in 
  panel d. 
   }
\end{figure}

In figure \ref{fig2}, the ICSs for electron S$_2$ collisions 
are shown for excitations to the ${A'}^3\Delta_{u}$, ${A}^3\Sigma_{u}^{+}$,
${B'}^3\Pi_{g}$ and ${B}^3\Sigma_{u}^{-}$ states. 
The ICS of Garrett et al. \cite{S2IP} obtained by the impact-parameter method is also 
compared with our excitation ICS to the ${B}^3\Sigma_{u}^{-}$ state.
In general, the slopes of these ICSs near threshold are steeper than those 
in fig.\ref{fig1}. 
The maximum values of the ICSs below 15 eV are about 0.3, 0.1, 0.6 and 
3.0 for the excitation to the ${A'}^3\Delta_{u}$, ${A}^3\Sigma_{u}^{+}$,
${B'}^3\Pi_{g}$ and ${B}^3\Sigma_{u}^{-}$ states, respectively. 
For the optically allowed ${B}^3\Sigma_{u}^{-}$ state excitation, 
the effect of the high $l$ partial waves is included by the Born closure approximation 
formula given in eq.(1). The magnitude of the Born correction is small 
below 10 eV, however, it increases as the scattering energy increases.  
The fraction of the correction to the R-matrix cross section becomes about 
20$\%$ at 15 eV. 
Our ICS with the Born correction and the previous result of Garrett et al. \cite{S2IP} 
agree reasonably well above 8 eV. We used the CASSCF value for the ${B}^3\Sigma_{u}^{-}$ 
excitation energy in this work, whereas Garrett et al. \cite{S2IP} employed the 
experimental value for the excitation energy. 
In contrast to the fixed-nuclei approximation in our calculation, 
Garrett et al. \cite{S2IP} included the effect of S$_2$ vibration in their calculation. 
Because of these differences, the results of Garrett et al. \cite{S2IP} and 
our ICS do not agree well near the excitation threshold. 
A small peak is seen at 14.2 eV in the ICSs of the ${A'}^3\Delta_{u}$ and 
${A}^3\Sigma_{u}^{+}$ state excitations, which is originated from the  
${}^4 \Pi_g$ symmetry partial cross sections. 
Since the location of this peak is higher than the highest energy 
S$_2$ electronic state included in the present R-matrix calculation, 
we cannot determine whether this peak belongs to the real resonance or 
pseudo-resonance. 

\begin{figure}
\includegraphics{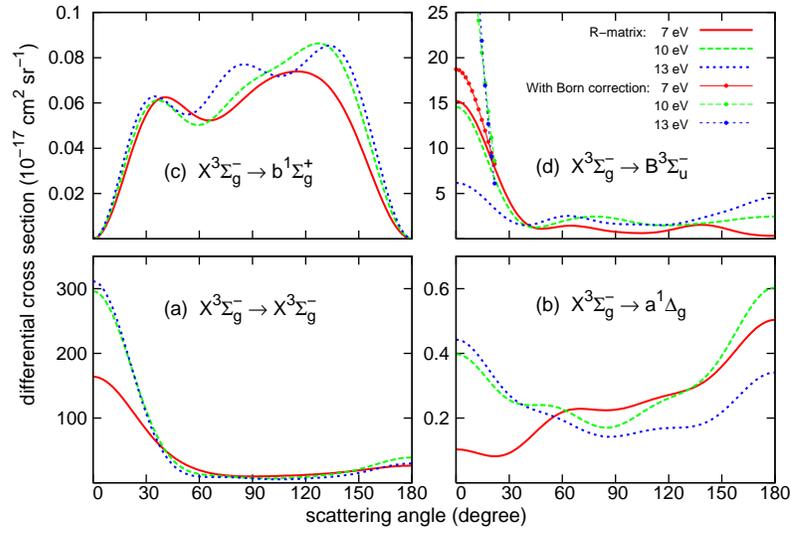}%
 \caption{\label{fig3} 
 Differential cross sections of elastic collision (panel a) and 
 excitation of the $a^{1} \Delta_{g}$ (panel b), 
 $b^{1} \Sigma_{g}^{+}$ (panel c) and 
 ${B}^{3} \Sigma_{u}^{-}$ (panel d) states.
 The R-matrix cross sections with Born correction are also included in 
 panel d. 
   }
\end{figure}

In figure\ref{fig3}, the differential cross sections (DCSs) are shown for 
elastic electron S$_2$ collisions as well as excitations to the 
${a}^1\Delta_{g}$, ${b}^1\Sigma_{g}^{+}$ and ${B}^3\Sigma_{u}^{-}$ states. 
The elastic DCSs are enhanced in forward direction and tend to be 
more forward-enhanced as the scattering energy increases. 
In contrast to the elastic DCSs, the excitation DCS to the 
${a}^1\Delta_{g}$ state has backward-enhanced character in general. 
However, the magnitude of forward scattering cross section 
increases as the scattering energy increases from 7 to 13 eV. 
Our excitation DCSs to the ${b}^1\Sigma_{g}^{+}$ state approach 
zero near 0 and 180 degrees, because of a selection rule associated with 
$\Sigma^{+}$-$\Sigma^{-}$ transition \cite{Go71,Ca71}. 
Around 90 degrees, the magnitude of the DCS is about 0.06$\sim$0.08 
$\times 10^{-17} {\rm cm}^2 {\rm sr}^{-1}$ and does not depend much on 
the scattering energy. 
The ${B}^3\Sigma_{u}^{-}$ state excitation DCSs are shown 
in fig. \ref{fig3} panel (d). In addition to the R-matrix results, 
the DCSs with the Born correction obtained by eq.(1) are also shown for scattering 
angles below 25 degrees. For larger angles, the magnitude of the correction is 
expected to be small and not shown here. 
The contribution of the Born correction to the DCS is small for 7 eV, however, 
it dominates the total DCS near zero degree at 10 and 13 eV. 
Although the magnitude of the R-matrix DCS at forward direction decreases as the 
scattering energy increases from 7 to 13 eV, the DCS with the Born correction at 
forward direction increases as energy increases. 


The elastic and excitation cross sections of the ${a}^1\Delta_{g}$ and 
${b}^1\Sigma_{g}^{+}$ states in electron S$_2$ collisions are about two times 
larger than corresponding cross sections in electron O$_2$ collisions 
studied in our previous paper \cite{tashiroO2ics}. 
Although the ${}^2 \Pi_u$ resonance peaks can be observed in the 
${a}^1\Delta_{g}$ and ${b}^1\Sigma_{g}^{+}$ excitation cross sections in 
both e-O$_2$ and e-S$_2$ collisions, the width of the peak is much broader 
in e-O$_2$ case. In e-O$_2$ elastic collision, a narrow ${}^2 \Pi_g$ 
resonance peak is seen below 1 eV. In e-S$_2$ elastic case, the energy of the 
S$_2^-$ ${}^2 \Pi_g$ state is stabilized below the energy of the S$_2$ ground 
state and cannot be observed in the cross section. 
Other than these resonance features, the profiles of the cross sections are 
similar in e-O$_2$ and e-S$_2$ collisions. 

In this work, we employed the fix-bond approximation for the R-matrix 
calculation. In our previous studies on electron impact excitations of 
O$_2$ and N$_2$ molecules\cite{tashiroO2ics,tashiroO2DCS,tashiroN2}, 
we also used the same fix-bond approximation and 
got good agreement with available experimental results, even if the positions 
of the potential curve minimum are different between the ground state and 
the excited state.  
Thus, the results of this study is also expected to be accurate enough. 
For more precise comparison of the excitation cross section of the 
${B}^3\Sigma_{u}^{-}$ state with the previous results of Garrett et al. 
\cite{S2IP}, inclusion of the vibrational effect may be necessary. 
Such kind of calculation is possible by the non-adiabatic R-matrix method 
or the adiabatic averaging of the T-matrix elements, and will be performed in future. 


We have carried out the R-matrix calculations with the maximum $l$ quantum number 4, 5 
and 6 to check convergence. Except for the ${B}^3\Sigma_{u}^{-}$ state excitation, 
the ICSs and DCSs are converged at $l$=5. For the ${B}^3\Sigma_{u}^{-}$ excitation, 
the ICS with the Born correction is also converged at $l$=5. 
For the ${B}^3\Sigma_{u}^{-}$ excitation DCSs with the Born correction, 
however, the convergence is achieved only below 25 degrees. Note that the similar 
situation was observed in Gibson et al.\cite{Gibson87}. 
Although the ${B}^3\Sigma_{u}^{-}$ excitation DCS is not converged above 30 degrees, 
the effect of the higher $l$ partial waves is expected to be small because 
the magnitude of the Born DCS itself is small at larger scattering angles. 
In principle, this convergence problem can be solved \cite{Rescigno1992,Sun1992} 
by applying the Born correction to T-matrix elements, 
where we applied the correction at the R-matrix DCS in this work. 
We will calculate this kind of Born correction at T-matrix level when 
more accurate excitation DCS is required in future. 

\section{Summary}
\label{4}
In this work, we have studied the low-energy electron impact excitations of 
S$_2$ molecules using the fixed-bond R-matrix method based on 
state-averaged CASSCF molecular orbitals. 
Thirteen target electronic states of S$_2$ are included 
in the model within a valence configuration interaction 
representations of the target states. 
Integral cross sections are calculated for elastic electron collision 
as well as impact excitation of the 7 lowest electronic states. 
Also, differential cross sections are shown for elastic collision 
and excitation of the ${a}^1\Delta_{g}$, ${b}^1\Sigma_{g}^{+}$ and 
${B}^3\Sigma_{u}^{-}$ states. 
For the excitations of the ${a}^1\Delta_{g}$ and ${b}^1\Sigma_{g}^{+}$ states, 
a narrow ${}^2 \Pi_u$ resonance peak is observed in the ICSs at 2.7 eV. 
For the elastic and the excitation collisions of the ${a}^1\Delta_{g}$ and 
${b}^1\Sigma_{g}^{+}$ states, the shapes of the cross sections are similar to 
those in electron O$_2$ collisions, however, the magnitudes of the cross 
sections are two time larger in electron S$_2$ collisions. 
Our ICS of the ${B}^3\Sigma_{u}^{-}$ state excitation agrees reasonably well  
with the previous result of Garrett et al. \cite{S2IP}. However, 
the ICS near threshold does not agree well, because of difference  
in excitation energy employed in calculation as well as the treatment of the 
vibrational effect. 

\clearpage




\begin{thebibliography}{00}


%

\bibitem{S2spectr}
G. Herzberg, Molecular Spectra and Molecular Structure, Vol. 1, 
Spectra of Diatomic Molecules, Van Nostrand Reinhold, New York, 1950. 
\bibitem{Swope79}
W. C. Swope, Y. P. Lee, H. F. Schaefer III, J. Chem. Phys. 70 (1979) 947. 
\bibitem{Saxon80}
R. P. Saxon, B. Liu, J. Chem. Phys. 73 (1980) 5174.

\bibitem{S2Jupiter}
K. S. Noll, M. A. McGrath, L. M. Trafton, S. K. Atreya, J. J. Caldwell, H.
A. Weaver, R. V. Yelle, C. Barnet, S. Edington, Science 267 (1995) 1307.

\bibitem{S2Io}
P. Geissler, A. McEwen, C. Porco, D. Strobel, J. Saur, J. Ajello, R. West, 
Icarus 172 (2004) 127. 

\bibitem{S2Hyakutake}
S. J. Kim, M. F. A'Hearn, D. D. Wellnitz, R. Meier, Y. S. Lee, 
Icarus 166 (2003) 157. 

\bibitem{S2etching}
L. St-Onge, N. Sadeghi, J. P. Booth, J. Margot, C. Barbeau, 
J. Appl. Phys. 78 (1995) 6957. 

\bibitem{S2Lamp}
C. W. Johnston, H. W. P. van der Heijden, A. Hartgers,  
K. Garloff, J. van Dijk, J. J. A. M. van der Mullen,
J. Phys. D: Appl. Phys. 37 (2004) 211.

\bibitem{S2IP}
B.C. Garrett, L. T. Redmon, C. W. McCurdy, M. J. Redmon, 
Phys. Rev. A 32 (1985) 3366. 

\bibitem{S2DAexp}
Y. Le Coat, L. Bouby, J. P. Guillotin, J. P. Ziesel, 
J. Phys. B: At. Mol. Opt. Phys. 29 (1996) 545. 

\bibitem{Tashiro2008}
M. Tashiro, Phys. Rev. A 77 (2008) in press. \\  
Preprint http://lanl.arxiv.org/abs/0712.1068v1

\bibitem{tashiroO2ics}
M. Tashiro, K. Morokuma, J. Tennyson, Phys. Rev. A 73 (2006) 052707.

\bibitem{tashiroO2DCS}
M. Tashiro, K. Morokuma, J. Tennyson, Phys. Rev. A 74 (2006) 022706.

\bibitem{tashiroN2}
M. Tashiro, K. Morokuma, Phys. Rev. A 75  (2007) 012720.

\bibitem{Mo98}
L. A. Morgan, J. Tennyson, C. J. Gillan, Comput. Phys. Commun. 114 (1998) 120.

\bibitem{Bu05}
P. G. Burke, J. Tennyson, Mol. Phys. 103 (2005) 2537.

\bibitem{Go05}
J. D. Gorfinkiel, A. Faure, S. Taioli, C. Piccarreta, G. Halmova, J. Tennyson, 
Eur. Phys. J. D 35 (2005) 231.

\bibitem{molpro}
MOLPRO, version 2006.1, a package of ab initio programs, H.-J. Werner, P. J. Knowles, R. Lindh, F. R. Manby, M. Sch{\"u}tz, and others, see http://www.molpro.net.


\bibitem{1993JChPh..98.1358W}
D. E. Woon, T. H. Dunning, Jr., J. Chem. Phys. 98 (1993) 1358.

\bibitem{Fa02}
A. Faure, J. D. Gorfinkiel, L. A. Morgan, J. Tennyson, Comput. Phys. Commun. 144 (2002) 224.

\bibitem{Crawford1971}
O. H. Crawford, A. Dalgarno, J. Phys. B: At. Mol. Opt. Phys. 4 (1987) 494. 

\bibitem{Gibson87}
T. L. Gibson, M. A. P. Lima, V. McKoy, W. M. Huo, Phys. Rev. A 35 (1987) 2473.

\bibitem{Itikawa1969}
Y. Itikawa, K. Takayanagi, J. Phys. Soc. Jpn. 26 (1969) 1254.

\bibitem{ChemPhys.71.79}
B. Hess, R. J. Buenker, C. M. Marian, S. D. Peyerimhoff, Chem. Phys. 71 (1982) 79. 

\bibitem{2000JChPh.112.7475K}
T. Kiljunen, J. Eloranta, H. Kunttu, L. Khriachtchev, M. Pettersson, and M. R{\"a}s{\"a}nen, J. Chem. Phys. 112 (2000) 7475.

\bibitem{Go71}
W. A. Goddard III, D. L. Huestis, D. C. Cartwright, S. Trajmar, Chem. Phys. Lett. 11 (1971) 329.

\bibitem{Ca71}
D. C. Cartwright, S. Trajmar, W. Williams, D. L. Huestis, Phys. Rev. Lett. 27 (1971) 704.

\bibitem{Rescigno1992}
T. N. Rescigno, B. I. Schneider, Phys. Rev. A 45 (1992) 2894.

\bibitem{Sun1992}
Q. Sun, C. Winstead, V. McKoy, Phys. Rev. A 46 (1992) 6987.






\end{thebibliography}

\clearpage

\begin{table}%
\caption{\label{tab1}
 The vertical excitation energies of the first 8 excited states 
 for S$_2$ molecule, with the previous MRD CI results of 
 Hess et al.\cite{ChemPhys.71.79}, 
 MRCI results of Kiljunen et al.\cite{2000JChPh.112.7475K} 
 and experimental values quoted in Hess et al.\cite{ChemPhys.71.79}.  
 The unit of energy is eV.
}
\begin{tabular}{ccccc}
\hline
State & This work & Previous MRD CI & Previous MRCI & Expt.\\
\hline
${X}^3 \Sigma_{g}^{-}$  &  0.00  & 0.00 & 0.00  & 0.00 \\
${a}^1 \Delta_{g}$      &  0.60  & 0.68 & 0.55  & 0.71 \\
${b}^1 \Sigma_{g}^{+}$  &  0.92  & 1.04 & 0.99  & 0.99 \\
${c}^1 \Sigma_{u}^{-}$  &  2.77  &      & 2.45  &      \\
${A'}^3 \Delta_{u}$     &  2.93  &      & 2.59  &      \\
${A}^3 \Sigma_{u}^{+}$  &  3.03  &      & 2.58  &      \\
${B'}^3 \Pi_{g}$        &  4.84  & 4.63 & 4.36  & 4.38 \\
${B}^3 \Sigma_{u}^{-}$  &  5.03  &      & 3.89  &      \\
\hline
\end{tabular}
\end{table}

\clearpage

\clearpage

\clearpage

\end{document}